\documentclass[useAMS, usenatbib]{mn2e}

\usepackage{graphicx}

\title[Comparing RV of atmospheric lines with radiosonde data]{Comparing radial velocities of atmospheric lines with radiosonde measurements}

\author[P. Figueira, F. Kerber , A. Chacon,  et al.]{P. Figueira$^{1}$\thanks{E-mail:
pedro.figueira@astro.up.pt}, F. Kerber$^{2}$, A Chacon$^{3}$, C. Lovis$^{4}$, N.C. Santos$^{1}$, G. Lo Curto$^{2}$, 
\newauthor M. Sarazin$^{2}$ and F. Pepe$^{4}$\\
$^{1}$Centro de Astrof\'{i}sica, Universidade do Porto, Rua das Estrelas, 4150-762 Porto, Portugal \\
$^{2}$European Southern Observatory, Karl-Schwarzschild-Strasse 2, D-85748 Garching bei M\"unchen, Germany \\
$^{3}$Universidad de Valpara\'{i}so, Av. Gran Breta\~{n}a 1111, Valpara\'{i}so, Chile \\
$^{4}$Observatoire Astronomique de l'Universit\'{e} de Gen\`{e}ve, 51 Ch. des Maillettes, 
    - Sauverny - CH1290, Versoix, Suisse \\
}
\begin{document}

\date{Accepted 2011 October 14.  Received 2011 October 10; in original form 2011 August 29}

\pagerange{\pageref{firstpage}--\pageref{lastpage}} \pubyear{2011}

\maketitle

\label{firstpage}

\begin{abstract}
The precision of radial velocity (RV) measurements depends on the precision attained on the wavelength calibration. One of the available options is using atmospheric lines as a natural, freely available wavelength reference. Figueira et al. (2010) measured the RV of O$_2$ lines using HARPS and showed that the scatter was only of $\sim$10\,m/s over a timescale of 6\,yr. Using a simple but physically motivated empirical model, they demonstrated a precision of 2\,m/s, roughly twice the average photon noise contribution. In this paper we take advantage of a unique opportunity to confirm
the sensitivity of the telluric absorption lines RV to different atmospheric and observing conditions: by means of contemporaneous in-situ wind measurements. This opportunity is a result of the work done during site testing and characterization for the European Extremely Large Telescope (E-ELT). The HARPS spectrograph was used to monitor telluric standards while contemporaneous atmospheric data was collected using radiosondes. We quantitatively compare the information recovered by the two independent approaches.

The RV model fitting yielded similar results to that of Figueira et al. (2010), with lower wind magnitude values and varied wind direction. The probes confirmed the average low wind magnitude and suggested that the average wind direction is a function of time as well. However, these results are affected by large uncertainty bars that probably result from a complex wind structure as a function of height. The two approaches deliver the same results in what concerns wind magnitude and agree on wind direction when fitting is done in segments of a couple of hours. Statistical tests show that the model provides a good description of the data on all timescales, being always preferable to not fitting any atmospheric variation. The smaller the timescale on which the fitting can be performed (down to a couple of hours), the better the description of the real physical parameters. We conclude then that the two methods deliver compatible results, down to better than 5\,m/s and less than twice the estimated photon noise contribution on O$_2$ lines RV measurement. However, we cannot rule out that parameters $\alpha$ and $\gamma$ (dependence on airmass and zero-point, respectively) have a dependence on time or exhibit some cross-talk with other parameters, an issue suggested by some of the results.

\end{abstract}

\begin{keywords}
Atmospheric effects, Instrumentation: spectrographs, Methods: observational, Techniques: radial velocities
\end{keywords}

\section{Introduction}

The research on extrasolar planets is currently one of the fastest-growing in Astrophysics. Triggered by the pioneering work of \cite{1995Natur.378..355M} on 51\,Peg, it evolved into a domain of its own, with more than 500 planets confirmed up to date. Most of these planets ($\sim$90\%) were detected using the radial velocity (RV) induced on the star by the orbital motion of the planet around it. 
The measurement of precise RVs can only be done against a precise wavelength reference, and two different approaches were pursued extensively. The first was the usage of a Th-Ar emission lamp with the cross-correlation function (CCF) method \citep{1996A&AS..119..373B},  and the second the I$_2$ cell along with the deconvolution procedure \citep{1996PASP..108..500B}.
In order to measure precise RV in the IR with CRIRES \citep{2004SPIE.5492.1218K},  \citet{2010A&A...511A..55F} recovered a method known for a long time: the usage of atmospheric features as a wavelength anchor. Using CO$_2$ lines present in the H band, the authors reached a precision of $\sim$5\,m/s over a timescale of one week. While a similar precision had been attained in the past in the optical domain using O$_2$ lines, the studies on the stability of atmospheric lines were limited to a timescale of up to a couple of weeks.

In order to assess the RV stability of atmospheric lines over longer timescales, \cite{2010A&A...515A.106F} used HARPS (High Accuracy Radial velocity Planet Searcher) archival data, spanning more than six years. Three stars  -- Tau Ceti, $\mu$ Arae, and $e$ Eri -- were selected because they provided a strong luminous background against which the atmospheric lines could be measured, and were observed not only over a long timespan but with high temporal frequency (in astereoseismology campaigns). The spectra were cross-correlated against an O$_2$ mask using HARPS pipeline, which delivered the RV, bisector span (BIS) and associated uncertainties. The high intrinsic stability of HARPS allowed one to measure these effects down to 1\,m/s of precision, roughly the photon noise attained on the atmospheric lines.

The r.m.s. of the velocities turned out to be of only $\sim$10\,m/s, and yet well in excess of the attained photon noise. An inspection of the RV pattern on one star over one night revealed not white noise but a well-defined shape on RV, BIS, contrast and FWHM. A component of the RV signal was associated with BIS variation, which in turn was linearly correlated with the airmass at which the observation was performed. A second component of the signal was interpreted as being the translation of the atmospheric lines' center created by the projection of an average horizontal wind vector along the line of sight. These two effects were described by the formula

\begin{equation}
	\Omega =  \alpha \times  {\left({1\over sin(\theta)} - 1\right)} + \beta \times cos(\theta) \,.\, cos(\phi -  \delta) + \gamma \label{eq_fit}
\end{equation}

where $\alpha$ is the proportionality constant associated with the variation in airmass, $\beta$ and $\delta$ the average wind speed magnitude and direction, and $\theta$ and $\phi$ the telescope elevation and azimuth, respectively. The $\gamma$ represents the zero-point of the RV, which can differ from zero. The fitting of the variables $\alpha$, $\beta$, $\gamma$, and $\delta$ allowed a good description of the telluric RV signal, with the scatter around the fit being of around 2\,m/s, or twice the photon noise. The fitting was performed in two ways: first, allowing all parameters to vary freely and second, imposing the same $\alpha$ and $\gamma$ for the different datasets. For details the interested reader is referred to the original paper. However, the model represented by Eq.\ref{eq_fit}, while being physically motivated, was not fully validated due to the absence of wind measurements against which the fitted values could be compared. 

Among the atmospheric parameters studied for E-ELT site testing is precipitable water vapor (PWV),  the major contributor to the opacity of Earth's atmosphere in the infrared. Hence the mean PWV established over long timescales determines how well a site is suited for IR astronomy. For the E-ELT site characterisation a combination of remote sensing (satellite data) and on-site data was used to derive the mean PWV for several potential sites, taking La Silla and Paranal as reference \citep{2010SPIE.7733E..48K}. In order to better understand the systematics in the archival data and to obtain data at higher time resolution, a total of three campaigns were conducted at La Silla Paranal observatory in 2009. During each campaign all available facility instruments as well as dedicated IR radiometers were used to measure PWV from the ground \citep[Kerber et al. {\it in prep.}][]{2011PASP..123..222Q}. In addition, radiosondes were launched to measure the vertical profile of atmospheric parameters in situ, with the goal of calculating the real PWV in the atmosphere. Radiosondes are an established standard in atmospheric research and all other methods were validated with respect to the radiosonde results with very high fidelity \citep{2010SPIE.7733E..48K, 2010SPIE.7733E.135Q, 2010SPIE.7733E.143C}.  

In the current paper we present the results of exploiting data from the above campaigns: since HARPS observations and radiosonde measurements were done in parallel we are in a position to make a direct and quantitative comparison of the wind speed parameters ($\beta$ and $\delta$). The paper is structured as follows. In Sect.\,2 we describe the data acquisition and reduction of both observing campaigns. Section\,3 is dedicated to the description of the analysis of data and subsequent results. In Sect.\,4 we discuss the implications of our results and we conclude in Sect.\,5 with the lessons learned from this campaign.

\section{Observations \& Data Reduction}

\subsection{HARPS measurements}

HARPS \citep{2003Msngr.114...20M} is a high-resolution fiber-fed cross-dispersed echelle spectrograph installed at the 3.6\,m telescope at La Silla Observatory. It is characterized by a spectral resolution of 110 000 and its 72 orders cover the the whole optical range, from 380 to 690\,nm. Its extremely high stability allows one to measure RV to a precision of better than 60\,cm/s when a simultaneous Th-Ar lamp is used, and of around 1\,m/s without the lamp. A dedicated pipeline (nicknamed DRS for {\it Data Reduction Software}) was created to allow for on-the-fly data reduction and RV calculation. This pipeline delivers the RV by cross-correlating the obtained spectra with a weighted binary mask. To calculate the atmospheric lines RV variation one needs then only to construct a template mask representing the lines to monitor. This  weighted binary mask \citep{2002A&A...388..632P} was built using HITRAN database \citep{2005JQSRT..96..139R} to select the O$_2$ lines present in HARPS wavelength domain. For the details on HARPS, the data reduction procedure and the mask construction, the reader is referred to \cite{2010A&A...515A.106F}. The procedure is identical, with the exception that the observations used in the current paper were performed without simultaneous Th-Ar.

For this program, 9 stars were observed: HR3090, HR3476, HR4748, HR5174, HR5987, HR6141, HR6930, HR7830, and HR8431, which are fast-rotating A-B stars, mostly featureless in the optical domain and suitable to be used as telluric standards. For details on the stars the reader is referred to the website ``Stars for Measuring PWV with MIKE"\footnote{http://www.lco.cl/operations/gmt-site-testing/stars-for-measuring-pwv-with-mike/stars-for-measuring-pwv} and to \cite{2007PASP..119..697T}. A total of 1120 measurements were collected on 8 and 9 of May, 2009, during the course of two nights of technical time. The stars were observed in a complex pattern in such a way that both low and high airmass and different patches of the sky were probed throughout the night in order to sample any variations of PWV. The main consequence is that even a fraction of the night with a couple of hours can contain observations of several stars at a wide range of airmass and elevation/azimuth coordinates, covering well the independent variables of Eq.\,\ref{eq_fit}. and allowing a precise estimation of the parameters to be fit.

\subsection{Radiosonde measurements}

The radiosonde (Vaisala RS-92) is a self-contained instrument package with sensors to measure e.g. temperature and humidity combined with a GPS receiver and a radio transmitter that relays all data in real-time to a receiver on the ground. The radiosonde is tied to a helium filled balloon and after launch ascends at a rate of a few m/s following the prevailing winds. On its ascent trajectory the sonde will sample the local atmospheric conditions up to an altitude of $\sim$20\,km, when the balloon will burst. By that time it has traveled horizontally $\sim$100 km from the launch site. Since it relays its 3D location based on GPS location every two seconds, the wind vector exerting force on the balloon can be deduced from the change in GPS position. 

A total of 17 radiosondes were launched between the 5 and the 15 of May of 2009 from La Silla site. One or two launches were conducted every day/night. On the 13th no data were collected due to a technical problem when radio contact with the radiosonde was lost shortly after lunch. From the collected physical parameters, the six of interest for our study, as well as the nominal precision of the measurements are presented in Tab. \ref{Table_sondes_prec}. As the sondes rise in height, they measure the two horizontal wind components on each layer with a nominal precision of 1$\times 10^{-3}$\,m/s, much higher than that of contemporaneous RV measurements.

Radiosondes form the backbone of the global network coordinated by the World Meteorological Organisation (WMO) for measuring conditions at the surface and in Earth's atmosphere by combining the in-situ atmospheric sounding with measurements taken onboard ships aircraft and satellites. Coordinated radiosonde launches (one launch at 12:00 UTC is the minimum requirement, other launch times are 00, 06 and 18 hours UTC) provide a global snapshot of atmospheric conditions which are then used as basic input for describing its current state and for modeling future conditions. 

The recommended maximum distance between stations is 250 km but the global distribution is very uneven and biased towards heavily populated areas in the Northern hemisphere. South America is sparsely covered with Chile operating 4 stations only one of which (Santo Domingo, WMO station number 85586) launches two radiosondes per day at 00 and 12 UTC. Data from all active launch sites can be found at http://weather.uwyo.edu/upperair/sounding.html.

The WMO also defines the requirements in terms of equipment and procedures such as number of barometric pressure levels, etc. A number of different radiosondes from different manufacturers are used in the various countries. To ensure comparability the WMO regularly conducts cross-calibration campaigns with parallel measurements (Jauhiainen \& Lehmuskero, 2005)\footnote{Performance of the Vaisala Radiosonde RS92-SGP and Vaisala DigiCORA Sounding System MW31 in the WMO Mauritius Radiosonde Intercomparison, February 2005. webpage http://www.vaisala.com/Vaisala\%20Documents/White\%20Papers/  Vaisala\%20Radiosonde\%20RS92\%20in\%20Mauritius\%20Intercomparison.pdf }. The Vailsala radiosonde RS-92 used in our campaign is considered to be the most reliable and accurate commercial device available. Its minor biases in particular for day-time launches are well-documented \citep[Jauhiainen \& Lehmuskero, 2005,][]{Milosh2009}. 

The global snapshot of the state of the atmosphere, taken at 00:00 and 12:00 UT is used as initial conditions of global meteorological numerical models (GFS(1), ECMWF(2), GME(3) among others\footnote{(1) Global Forecast System (http://www.emc.ncep.noaa.gov/gmb/ moorthi/gam.html) 

(2) European Center of Medium range Weather Forecasting (http://www.ecmwf.int/products/data/operational\_system/description/ brief\_history.html) 

(3) Global Numerical Weather Prediction Model (http://journals.ametsoc.org/doi/abs/10.1175/1520-0493(2002) 130\%3C0319\%3ATOGIHG\%3E2.0.CO\%3B2) 

(4) The Fifth-Generation NCAR/ Penn State Mesoscale Model. NCAR=National Center of Atmospheric Research (http://www.mmm.ucar.edu/mm5/) 

(5) Weather Research and Forecasting model (http://www.wrf-model.org/index.php) 

(6) Non-Hydrostatic Mesoscale Atmospheric Model (http://mesonh.aero.obs-mip.fr/mesonh/)}).

These initial conditions are employed in numerical approximations using dynamical equations, which predict the future state of the atmospheric circulation \citep{Holton2005}. The models are a simplification of the atmosphere because the horizontal resolution of the grid can be between 60 km (GME) to 100 km (GFS) - and sometimes more - and the vertical resolution provides only very small number of layers, but on a global scale the results are very good and have improved considerably over the past decades. There are other models called mesoscales models (MM5 (4), WRF(5), MesoNH(6), among others), which provide higher spatial resolution (horizontal and vertical) use more specific dynamical equations (physics parameterization) and better resolution of the surface terrain. The initial conditions for these models are usually the global model augmented by some local weather stations. Details on the models are available from the sites mentioned above. 

Concerning the applicability of RS data to our purpose it is important to note that the radiosonde is the accepted standard in atmospheric and meteorological research. For global weather forecasting a distance of order 250 km between radiosonde launch stations is the desired but by no means always achieved standard, while the cadence is between 6 and 24 h. Hence, the radiosonde data set that we use for comparison with HARPS observations is well within the accepted limits of applicability in terms of spatial and time resolution. 

It is evident that local topography and diurnal variations may limit the value of a set of radiosonde data to smaller distances and shorter periods of time. To this end there is a very instructive analysis by \cite{Norbert-Kalthoff:2002cr} that is directly applicable to our case. They use the Karlsruhe Atmospheric mesoscale model (KAMM) and compare with wind measurements taken at stations around 30 degrees South in Chile, including the Cerro Tololo Interamerican Observatory (CTIO). La Silla (70$^{\circ}$44'4"5 W 29$^{\circ}$15'15"4 S) is located within that region only about 100 km N of CTIO. \cite{Norbert-Kalthoff:2002cr} find that the wind patterns over this region are stable, their diurnal variations are highly reproducible and that wind conditions are mostly stable during night time. Their main finding is that for altitudes between 2 and 4 km northerly winds prevail whereas above 4 km large scale westerly winds dominate. The reason for the Northerly wind is a deflection of westerly winds by the Cordillera de los Andes which forms a barrier. They provide a physical explanation (their section 4) in terms of the Froude number (ratio between inertial forces and buoyancy) demonstrating that this deflected northerly flow is a naturally stable phenomenon. As mentioned La Silla is located in the same region and the wind roses of Cerro Tololo (2200 m) 
(their Fig. 7) and La Silla (2400m)\footnote{http://www.eso.org/gen-fac/pubs/astclim/lasilla/humidity/ LSO\_meteo\_stat-2002-2006.pdf} are very similar, clearly showing a predominance of northerly wind. 

In particular winter months (June-August) are characterized by very constant daily ground wind properties (Fig 5 of the same paper). Our observations were made in May. 
In addition it is a well-established fact that wind conditions in the free atmosphere are much more stable than in the turbulent and highly variable ground layer \citep[see e.g.][]{Holton2005, Wallace2006}. 

As a consequence of the very homogeneous overall wind structure between 2 km and 4 km and above 4 km we have reason to believe, that the information on the wind vectors obtained by a radiosonde will be representative of conditions over the time span of at least a good fraction of a night for our campaigns. 

\begin{table*}
\centering
\caption{The six parameters used from the data collected by radiosondes.}
\begin{tabular}{lccc} \hline\hline
 \ \ Parameter & Unit & Precision [Unit] & Comment \\ \hline \hline
time  & s   & 0.1 & measurements cadence of one every 2\,s \\
T & K & 0.1&  --- \\
P & hPa & 0.1 & --- \\
Height & m & 1 & limited to 30\,km \\ 
u & m/s & 0.01 & E-W wind component, East positive\\
v & m/s & 0.01 & N-S wind component, North positive\\

\hline \hline
\end{tabular}

\label{Table_sondes_prec}
\end{table*}

\section{Analysis \& Results}

\subsection{HARPS measurements}

We analyzed the data from the 9 stars as if coming from a single data set, as there is no reason to treat them separately. As done in \cite{2010A&A...515A.106F}, we discarded the 27 datapoints with photon noise precision worse than 5\,m/s, which correspond to only 2.5\% of the observations.

The total RV scatter and average photon noise were 5.01\,m/s and 2.82\,m/s, respectively. If one separates the set in the two nights that constitute it, the values for the first night are of 5.36 and 2.92\,m/s,  and 4.60 and 2.72\,m/s for the second night. We note that the photon noise contribution to the precision from the stellar spectrum is larger than 1\,m/s, validating the choice of not using the lamp simultaneously with the observations.

We fitted the RV variation on the two nights using Eq.\,\ref{eq_fit}, as described in  \cite{2010A&A...515A.106F}. When fitting, we considered splitting the dataset in three different ways and making two different hypothesis for the parameters variation. On the splitting of the dataset we employed: 1) the same parameters for all the observations, 2) an independent set of parameters per night, and 3) a set of parameters for each one-third of the night. After allowing all parameters to vary freely at first, we repeated this imposing $\alpha$ and $\gamma$ to be the same for the the whole dataset in 2) and 3). The resulting parameters, $\chi^2_{red}$, and scatter around the fit are presented in Tab. \ref{fitstats} for each case. The error bars were estimated by bootstrapping the residuals and repeating the fitting 10000 times. The 95\% confidence intervals were drawn from the distribution of the parameters, and the 1$\sigma$ uncertainty estimations are presented.

\begin{table*}

\caption{The fitted parameters and data properties, before and after the fitted model is subtracted from it, with all parameters kept varying freely ({\it top}), and when $\alpha$ and $\beta$ are imposed as the same for the datasets ({\it bottom}, marked with {\it *}).}\label{fitstats}

\begin{tabular}{lccccccccc} \hline\hline

 \ \  data set  &  $\#$obs  & $\sigma$\,[m/s] &  $\sigma_{(O-C)}$\,[m/s]  &  $\sigma_{ph}$\,[m/s] & $\chi^{2}_{red}$  & $\alpha$\,[m/s] &  $\beta$\,[m/s]  & $\gamma$\,[m/s] &  $\delta$\,[$^{o}$] \\[1pt] \hline \hline
08+09-05-2009 & 1093 &   5.01 &   4.14 &    2.82  &   2.20 & 7.79$_{-0.73}^{+0.76}$ & 8.47$_{-0.68}^{+0.76}$ & 220.56$_{-0.31}^{+0.05}$ & 126.10$_{-5.56}^{+4.41}$  \\[2pt]
\hline
 08-05-2009 (1$^{st}$ n.) & 554 &   5.36 &   4.38 &    2.92  &   2.12 & 5.74$_{-1.81}^{+1.87}$ & 13.93$_{-2.61}^{+3.12}$ & 220.32$_{-0.79}^{+0.31}$ & 145.14$_{-13.00}^{+5.58}$ \\[4pt]
 09-05-2009 ($2^{nd}$ n.) & 539 &   4.60 &   3.68 &    2.72  &   1.83 & 8.20$_{-0.81}^{+0.83}$ & 6.75$_{-0.50}^{+0.60}$ & 219.95$_{-0.04}^{+0.36}$ & 97.75$_{-7.91}^{+6.67}$  \\[2pt]
 
 \hline
 
 (1$^{st}$ n., section\,1/3) & 185 &   3.46 &   2.89 &    1.88  &   2.25  & 8.97$_{-1.41}^{+1.41}$ & 4.44$_{-0.89}^{+3.31}$ & 223.58$_{-0.75}^{+0.97}$ & 42.17$_{-17.60}^{+47.29}$ \\[4pt]
 
 (1$^{st}$ n., section\,2/3) & 185 &   4.58 &   4.58 &    3.27  &   1.71  & 10.79$_{-17.98}^{+21.27}$ & 15.99$_{-8.00}^{+43.88}$ & 223.96$_{-4.91}^{+5.24}$ & 17.22$_{-96.66}^{+77.04}$  \\[4pt]

 (1$^{st}$ n., section\,3/3) & 184 &   6.26 &   4.51 &    3.60  &   1.53  & 29.77$_{-12.13}^{+11.86}$ & 74.86$_{-31.07}^{+41.90}$ & 230.09$_{-5.30}^{+5.25}$ & 4.28$_{-2.35}^{+63.12}$ \\[4pt]

 (2$^{nd}$ n., section\,1/3) & 180 &   4.80 &   3.30 &    2.52  &   1.73  & 15.50$_{-1.71}^{+1.70}$ & 15.98$_{-3.19}^{+4.50}$ & 223.51$_{-1.18}^{+1.28}$ & 24.96$_{-7.08}^{+18.08}$ \\[4pt]

 (2$^{nd}$ n., section\,2/3) & 180 &   4.50 &   3.74 &    2.84  &   1.75  & 8.76$_{-2.38}^{+2.47}$ & 3.61$_{-0.78}^{+5.17}$ & 220.10$_{-0.50}^{+0.91}$ & 90.87$_{-47.05}^{+31.05}$  \\[4pt]

 (2$^{nd}$ n., section\,3/3) & 180 &   3.86 &   3.48 &    2.82  &   1.58  & -15.06$_{-18.04}^{+18.00}$ & 7.78$_{-1.77}^{+2.77}$ & 220.61$_{-0.70}^{+0.74}$ & 66.26$_{-27.05}^{+17.98}$  \\[2pt]
 
 \hline \hline

 
 08-05-2009 (1$^{st}$ n.)* & 554 &   5.36 &   4.38 &   2.92 &   2.12$^\dagger$ &   6.85 $_{-  0.77}^{+  0.76}$ &    13.57 $_{-  1.25}^{+  1.31}$ &   220.19 $_{-  0.23}^{+  0.13}$ &   143.17 $_{-  3.72}^{+  2.94}$ \\[4pt]
 09-05-2009 ($2^{nd}$ n.)* & 539 &   4.60 &   3.68 &   2.72 &   1.83$^\dagger$ &   6.85 $_{-  0.77}^{+  0.76}$ &     6.95 $_{-  0.60}^{+  0.69}$ &   220.19 $_{-  0.23}^{+  0.13}$ &   102.14 $_{-  8.25}^{+  6.81}$ \\[4pt]
 global fit parameters & 1093 & 5.01 &   4.06 &   2.82 &   1.98 & --- & --- & --- & --- \\[2pt]
 
 \hline
 
 (1$^{st}$ n., section\,1/3)* & 185 &   3.46 &   2.97 &   1.88 &   2.34$^\dagger$ &    8.81 $_{-  1.09}^{+  1.09}$ &     6.05 $_{-  1.41}^{+  2.00}$ &   221.39 $_{-  0.39}^{+  0.42}$ &   127.53 $_{- 23.16}^{+ 10.59}$ \\[4pt]
 
 (1$^{st}$ n., section\,2/3)* & 185 &   4.58 &   4.62 &   3.27 &   1.73$^\dagger$ &   8.81 $_{-  1.09}^{+  1.09}$ &    14.38 $_{-  4.70}^{+  8.65}$ &   221.39 $_{-  0.39}^{+  0.42}$ &   131.27 $_{- 53.92}^{+  7.37}$  \\[4pt]

 (1$^{st}$ n., section\,3/3)* & 184 &   6.26 &   4.55 &   3.60 &   1.55$^\dagger$ &   8.81 $_{-  1.09}^{+  1.09}$ &    11.17 $_{-  0.55}^{+  1.44}$ &   221.39 $_{-  0.39}^{+  0.42}$ &    77.45 $_{- 15.42}^{+ 16.67}$ \\[4pt]

 (2$^{nd}$ n., section\,1/3)* & 180 &   4.80 &   3.49 &   2.52 &   1.90$^\dagger$ &   8.81 $_{-  1.09}^{+  1.09}$ &     8.00 $_{-  0.73}^{+  1.01}$ &   221.39 $_{-  0.39}^{+  0.42}$ &    71.89 $_{- 13.51}^{+ 14.36}$ \\[4pt]

 (2$^{nd}$ n., section\,2/3)* & 180 &   4.50 &  3.78 &   2.84 &   1.78$^\dagger$ &   8.81 $_{-  1.09}^{+  1.09}$ &     9.23 $_{-  2.92}^{+  4.15}$ &   221.39 $_{-  0.39}^{+  0.42}$ &     5.39 $_{-  7.01}^{+ 36.17}$  \\[4pt]

 (2$^{nd}$ n., section\,3/3)* & 180 &   3.86 &  3.53 &   2.82 &   1.62$^\dagger$ &   8.81 $_{-  1.09}^{+  1.09}$ &    13.47 $_{-  1.59}^{+  1.82}$ &   221.39 $_{-  0.39}^{+  0.42}$ &    95.93 $_{-  9.63}^{+  8.41}$ \\[4pt]

 global fit parameters & 1093 &   5.01 &   3.87 &   2.82 &  1.81 & --- & --- & --- & --- \\[2pt]

\hline \hline

\end{tabular}

\medskip

Note that $\delta$\,=\,0\,$^{o}$ when wind direction points towards North, and positive eastwards. The error bars on each of the fitted parameters were drawn by bootstrapping the residuals (see text for details). Note that the $\chi^{2}_{red}$ marked with $^\dagger$ are not defined in the strict sense: they are calculated assuming 4 fitting parameters for the considered subset, with the objective of allowing comparison with the corresponding unconstrained fitting.

\end{table*}

While one might be tempted to compare the $\chi^2_{red}$ of the data as a way of quantifying the quality of the fit, there are several reasons not to do so. The first is that as one divides the data into subsets that are fitted independently, there is some ambiguity in how the $\chi^{2}_{red}$ of a set is compared with the combined $\chi^{2}_{red}$ of the subsets. However, more important is that we are considering a problem with {\it priors}, as the reader will realize when noting that $\beta \in$\, [0, $\infty$[ . The consequence is that this corresponds to the fitting of a non-linear model, for which the number of degrees of freedom is ill-defined, as recently underlined by \cite{2010arXiv1012.3754A}. In order to compare the quality of the data description by the different models, we follow the recommendations of the same authors. We calculate the probability that the normalized residuals of the fitting are drawn from a gaussian distribution with $\mu$\,=\,0 and $\sigma$\,=\,1, as expected if no signal is present and the scatter is dominated by the measurement uncertainty. To do so we use the Kolmogorov-Smirnov test \citep[as implemented in][]{1992nrfa.book.....P} and compute the probability $P_{KS}$  which, loosely speaking, corresponds to the probability that the residuals after fitting the model are drawn from a gaussian distribution. The larger the value of $P_{KS}$, the more appropriate the model is to describe the data-set in hands. We also calculated the probability $P_{KS}(no fit)$ for normalized RVs of each dataset without fitting the model, but to which only the average value was subtracted (which corresponds to fitting only a constant). The probability for each case on each data set is presented in Tab.\,\ref{KS_prob}.

\begin{table*}
\centering
\begin{tabular}{lccc} \hline\hline
 \ \ data set & P$_{KS}$ & P$_{KS}$(const. fit)  &  P$_{KS}$(no fit)  \\ \hline \hline

08+09-05-2009  & 1.47e-12      &       ---    &       4.32e-25 \\ 

08-05-2009 (1$^{st}$ n.)      & 2.10e-06      &       8.62e-06      &       6.28e-19 \\ 

 09-05-2009  ($2^{nd}$ n.)  & 9.29e-05      &       4.44e-04         &       2.62e-10 \\ 

 (1$^{st}$ n., section\,1/3)    & 1.36e-01      &      4.01e-03      &       1.22e-05 \\ 

 (1$^{st}$ n., section\,2/3)    & 4.68e-02      &       5.06e-02      &       3.02e-02 \\ 

 (1$^{st}$ n., section\,3/3)   & 1.46e-01      &        4.22e-02      &       1.07e-05 \\ 

 (2$^{nd}$ n., section\,1/3)   & 1.81e-01      &       3.31e-02      &       5.79e-08 \\ 

 (2$^{nd}$ n., section\,2/3)    & 1.22e-01      &       3.41e-01      &       4.66e-03 \\ 

 (2$^{nd}$ n., section\,3/3)     & 1.84e-01      &      7.93e-02      &       5.33e-02 \\ 

\hline \hline
\end{tabular}

\caption{The probability that the residuals and original datasets are drawn from gaussian distributions, as estimated used Kolmogorov-Smirnov test (see text for details). We represent by P$_{KS}$ the datasets in which the parameters can vary freely and P$_{KS}$(const. fit) the fit in which $\alpha$ and $\beta$ are imposed to be the same.} \label{KS_prob}
\end{table*}

\subsection{Radiosonde measurements }

The measurement of the radiosonde wind vector ($u,v$) as a function of time, or height, while interesting, is hardly insightful for our objective. We need to calculate the effect of this wind as integrated along the line of sight, such as it is measured by any telescope and spectrograph on the ground. This will deliver an average wind vector which can then be compared with the one obtained with HARPS (see the previous section).

A way of calculating this average wind is to consider a plane-parallel atmosphere that is composed of horizontal layers. Every radiosonde measurement probes the properties of a layer in its ascent. To obtain the average wind speed we weight the wind speed of each of these layers with its absorptivity. In doing so we are considering that the absorption line we measure with our spectrograph is the result of the product of the transmission of all layers, and each one of these creates a small line shifted by its respective horizontal wind. It is important to note that we chose doing so because absorptivity is proportional to the depth of the line at the central wavelength, and thus proportional to the spectral information contribution for the CCF as described in \cite{2002A&A...388..632P}.

The absorptivity on each layer A$_i$ is $A_i$ = 1 - e$^{-\tau}$ where $\tau$ is the optical depth and calculated as $\tau = I(T) \times Amplitude_{Lorentz} \times \sigma_{\mathrm{O}_2}(T,P) \times \Delta h$ where $I$ is the spectral line intensity, $Amplitude_{Lorentz}$ the relative amplitude of a Lorentzian function, $\sigma_{\mathrm{O}_2}$ the surface density of O$_2$, and $\Delta h$ the height of the layer in question.

The first component of $\tau$ is $I$, the spectral line intensity (basically, the line area) and is given in [cm$^{-1}$ / (molecule.cm$^{-2}$)] in HITRAN. Since $I$ is a function of $T$ we calculated a grid of HITRAN $I$ from the minimum to the maximum temperature measured by the radiosondes, with a step of 0.1\,K for all the O$_2$ lines within HARPS wavelength domain. For each temperature an average $I$ was assigned to the overall spectrum. This gives us $I(T)$, and to obtain values for T in between two grid points we fitted second-degree polynomials, which provided a very smooth description of the data. Interpolating the values provided the same wind values down to 0.01\,m/s.

In order to derive the line depth, one has to apply a correction to get the amplitude of the Lorentzian function that has the equivalent area, given by 1.0/($\pi \times HWHM$). The HWHM was set to 1.0, but its absolute value does not affect the results significantly, for it affects all layers in the same way. Subsequent tests showed that changing it from 0.1 to 10 led to variations of the order of 0.01\,m/s and 0.01\,$^{\circ}$ on wind magnitude and direction, much smaller than the error bars of the measurements.
The surface density $\sigma_{\mathrm{O}_2}(T,P)$ was calculated using the ideal gas law and assuming a constant volume mixing ratio (VMR) of O$_2$, of 20.946\,\% as function of height, which is a reasonable assumption up to 80\,km, hence well justified in the range of interest of up to 30\,km.

\begin{figure}
\includegraphics[width=9cm]{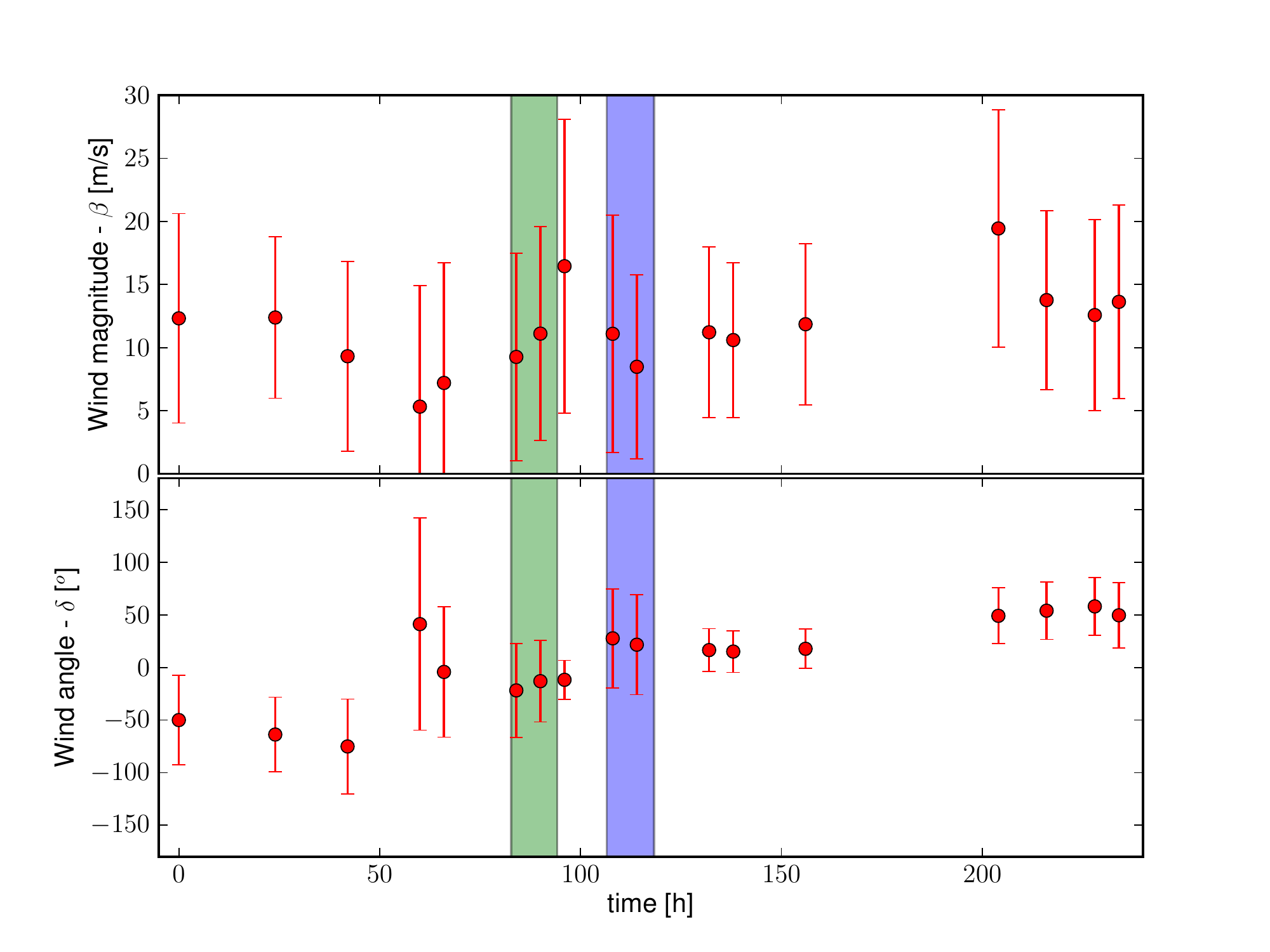}
\caption{Evolution of average wind magnitude ({\it upper panel}) and direction ({\it lower panel}), as measured by the radiosondes. The two nights during which observations with HARPS were performed are represented by shadowed, colored zones. The values are presented in Tab. \ref{Table_radiosondes}.}\label{windplot}
\end{figure}

With this we calculated the weighted average of the velocities, and a weighted standard deviation as well. The error on the average is estimated as being the weighted standard deviation. To allow comparison with the results from the previous section, one can calculate the vector magnitude and direction. The results are presented in Tab. \ref{Table_radiosondes} and plotted in Fig. \ref{windplot}.

\begin{table*}
\caption{Weighted average of wind components $u$ and $v$, as  well as wind vector magnitude and direction for each of the radiosondes launch.}\label{Table_radiosondes}

\begin{tabular}{cccccc} \hline\hline
\ \  \# probe & Observation date and hour  & $u$\,[m/s]  & $v$\,[m/s] & $\|u + v\|$\,[m/s] & $\delta\,[^{\circ}]$ \\ \hline \hline
 1 & EDT  /   05  /   05  /   09  /   1200\,UTC    &     $-$9.43 $\pm$ 5.71    &    7.91 $\pm$ 10.95    &    12.31 $\pm$ 8.28     &    $-$50.01 $\pm$ 42.62 \\
 2 & EDT  /   06  /   05  /   09  /   1200\,UTC    &     $-$11.10 $\pm$ 5.92    &    5.48 $\pm$ 8.05    &    12.38 $\pm$ 6.39     &    $-$63.71 $\pm$ 35.54 \\
 3 & EDT  /   07  /   05  /   09  /   0600\,UTC    &     $-$9.00 $\pm$ 7.53    &    2.40 $\pm$ 7.30    &    9.31 $\pm$ 7.52     &    $-$75.07 $\pm$ 45.00 \\
 4 & EDT  /   08  /   05  /   09  /   0000\,UTC    &     3.44 $\pm$ 8.55    &    4.09 $\pm$ 10.27    &    5.35 $\pm$ 9.60     &    40.03 $\pm$ 99.69 \\
 5 & EDT  /   08  /   05  /   09  /   0600\,UTC    &     $-$0.58 $\pm$ 7.76    &    7.22 $\pm$ 9.54    &    7.25 $\pm$ 9.53     &    $-$4.58 $\pm$ 61.42 \\
 6 &EDT  /   09  /   05  /   09  /   0000\,UTC    &     $-$3.43 $\pm$ 6.96    &    8.61 $\pm$ 8.40    &    9.27 $\pm$ 8.22     &    $-$21.70 $\pm$ 44.49 \\
 7 & EDT  /   09  /   05  /   09  /   0600\,UTC    &     $-$2.50 $\pm$ 7.46    &    10.85 $\pm$ 8.55    &    11.13 $\pm$ 8.50     &    $-$12.97 $\pm$ 38.69 \\ 
 8 &EDT  /   09  /   05  /   09  /   1200\,UTC    &     $-$3.34 $\pm$ 4.82    &    16.05 $\pm$ 11.78    &    16.40 $\pm$ 11.57     &    $-$11.75 $\pm$ 18.49 \\ 
 9 & EDT  /   10  /   05  /   09  /   0000\,UTC    &     5.16 $\pm$ 8.98    &    9.81 $\pm$ 9.48    &    11.09 $\pm$ 9.37     &    27.74 $\pm$ 46.99 \\
 10 & EDT  /   10  /   05  /   09  /   0600\,UTC    &     3.12 $\pm$ 6.95    &    7.88 $\pm$ 7.33    &    8.48 $\pm$ 7.28     &    21.58 $\pm$ 47.31 \\
 11 & EDT  /   11  /   05  /   09  /   0000\,UTC    &     3.24 $\pm$ 3.54    &    10.67 $\pm$ 7.00    &    11.15 $\pm$ 6.78     &    16.89 $\pm$ 20.30 \\ 
 12 & EDT  /   11  /   05  /   09  /   0600\,UTC    &     2.76 $\pm$ 3.34    &    10.20 $\pm$ 6.31    &    10.57 $\pm$ 6.16     &    15.14 $\pm$ 19.63 \\
 13 & EDT  /   12  /   05  /   09  /   0000\,UTC    &     3.63 $\pm$ 3.46    &    11.27 $\pm$ 6.62    &    11.84 $\pm$ 6.39     &    17.83 $\pm$ 18.70 \\
 14 & EDT  /   14  /   05  /   09  /   0000\,UTC    &     14.69 $\pm$ 10.42    &    12.54 $\pm$ 7.69    &    19.31 $\pm$ 9.37     &    49.53 $\pm$ 26.52 \\
 15 & EDT  /   14  /   05  /   09  /   1200\,UTC    &     11.00 $\pm$ 7.69    &    8.09 $\pm$ 5.96    &    13.66 $\pm$ 7.13     &    53.69 $\pm$ 27.77 \\
 16 & EDT  /   15  /   05  /   09  /   0000\,UTC    &     10.66 $\pm$ 8.44    &    6.66 $\pm$ 4.69    &    12.57 $\pm$ 7.57     &    57.98 $\pm$ 27.27 \\
 17 & EDT  /   15  /   05  /   09  /   0600\,UTC    &     10.32 $\pm$ 8.41    &    8.84 $\pm$ 6.49    &    13.59 $\pm$ 7.66     &    49.43 $\pm$ 31.05 \\
\hline \hline
\end{tabular}

\medskip

Note that $\delta$\,=\,0\,$^{o}$ when wind direction points towards North, and positive eastwards.

\end{table*}

\section{Discussion}

\subsection{The RV data}

The first point to note when it comes to the results of Tab.\,\ref{fitstats} is the low value of the r.m.s. of RVs over the two nights: 5\,m/s, less than twice the average photon noise value. As one selects smaller sets of data, first individual nights and then subsets of these nights, one obtains different fitted parameters. This suggests that the parameters are variable on a timescale smaller than one night. The higher probability $P_{KS}$ presented by the short-timescales datasets, attests to the fact that the model is more suitable to test the variability on smaller timescales. The fitting performed imposing the same $\alpha$ and $\gamma$ provides results with similar quality, better for the complete nights fitting, poorer if the nights are divided into subsections. Moreover, when comparing the fitted data with the raw data, one concludes that fitting the data with the models leads to residuals that are closer to gaussian than subtracting a constant from it. This means that the model description, even when less precise, still describes a fraction of the signal contained in the data and is always preferable over using raw data.

It is insightful to compare the data with that of \cite{2010A&A...515A.106F}. The fitted $\alpha$ and $\beta$ are comparable, and tend to be even lower, while $\gamma$ is similar in value. Importantly, $\delta$ varies significantly from one data-set to the other, just like it varied between the data sets considered in the previous paper. It is important to note that the error bars on some measurements are rather large, and the discrepancy between some consecutive measurements can be explained by this alone. In particular the last two thirds of the first night were affected by a high photon noise contribution, and these two sets yield the fits with the largest and more asymmetric error bars and largest residuals for the unconstrained fit. However, the scatter is already smaller than twice the photon noise, with or without subtracting the fit.

\begin{figure}
\includegraphics[width=9cm]{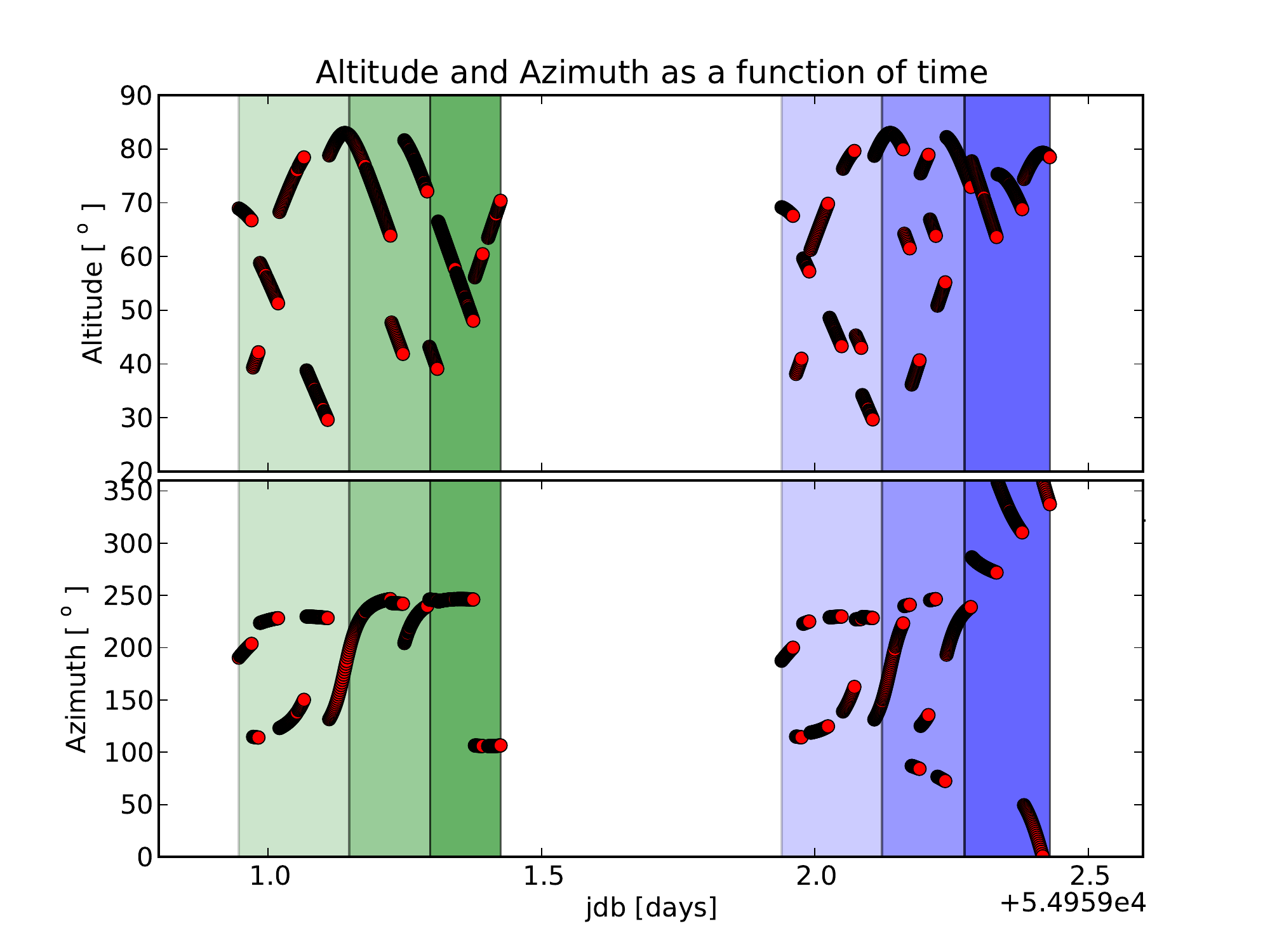}
\caption{The distribution of elevation and azimuth ($\theta$, $\phi$) as function of time, for the two observation nights. The three slices used for independent fitting are identified as shadowed regions with different colors.}\label{altaz}
\end{figure}

While the RV data were obtained with a different scientific objective than the one presented here (that of determining the PWV content in the atmosphere), the observations were still done in order to sample as many different patches of the sky as possible. However, it is extremely difficult to find suitable stars in all directions and thus sample evenly ($\theta$, $\phi$), our independent variables. We succeeded in obtaining observations for elevation $\theta \in$ [30, 85] $^o$ for both nights, but most observations were taken between azimuth angles $\phi \in$ [100, 250] $^o$, which might limit the accuracy with which the wind direction can be determined.  When one looks at the distribution of ($\theta$, $\phi$) as a function of time (Fig.\,\ref{altaz}) one concludes that the uneven distribution of the parameters to fit can limit the performance of the fit.

However, the main question that remains is if the model can be used imposing the same $\alpha$ and $\gamma$ for a given set of observations. While the results for the constrained fit are slightly worse, they do not allow us to reach a firm conclusion with regards to this aspect. 

\subsection{The radiosondes data}

The data from the radiosondes provide some interesting clues on the behavior of the atmosphere. The calculation of the average horizontal wind was affected by large uncertainties, a consequence of the complex and variable structure of winds as one travels across the atmosphere. However, the average wind magnitude is remarkably low, being between 5 and 16\,m/s. The values close in time are in agreement within error bars, both in what concerns magnitude and direction. Note, in particular, the values obtained using probes \#6,7, and 8, released 6 hours apart and perfectly compatible within their assigned uncertainty. It is interesting to point out that the measurement yielding the largest uncertainty on wind direction is the one with the smallest wind magnitude value, as expected. The wind direction measurements suggest also a slow variation of wind direction as a function of time.

\subsection{Comparing RV and radiosondes data}

\begin{figure}
\includegraphics[width=9cm]{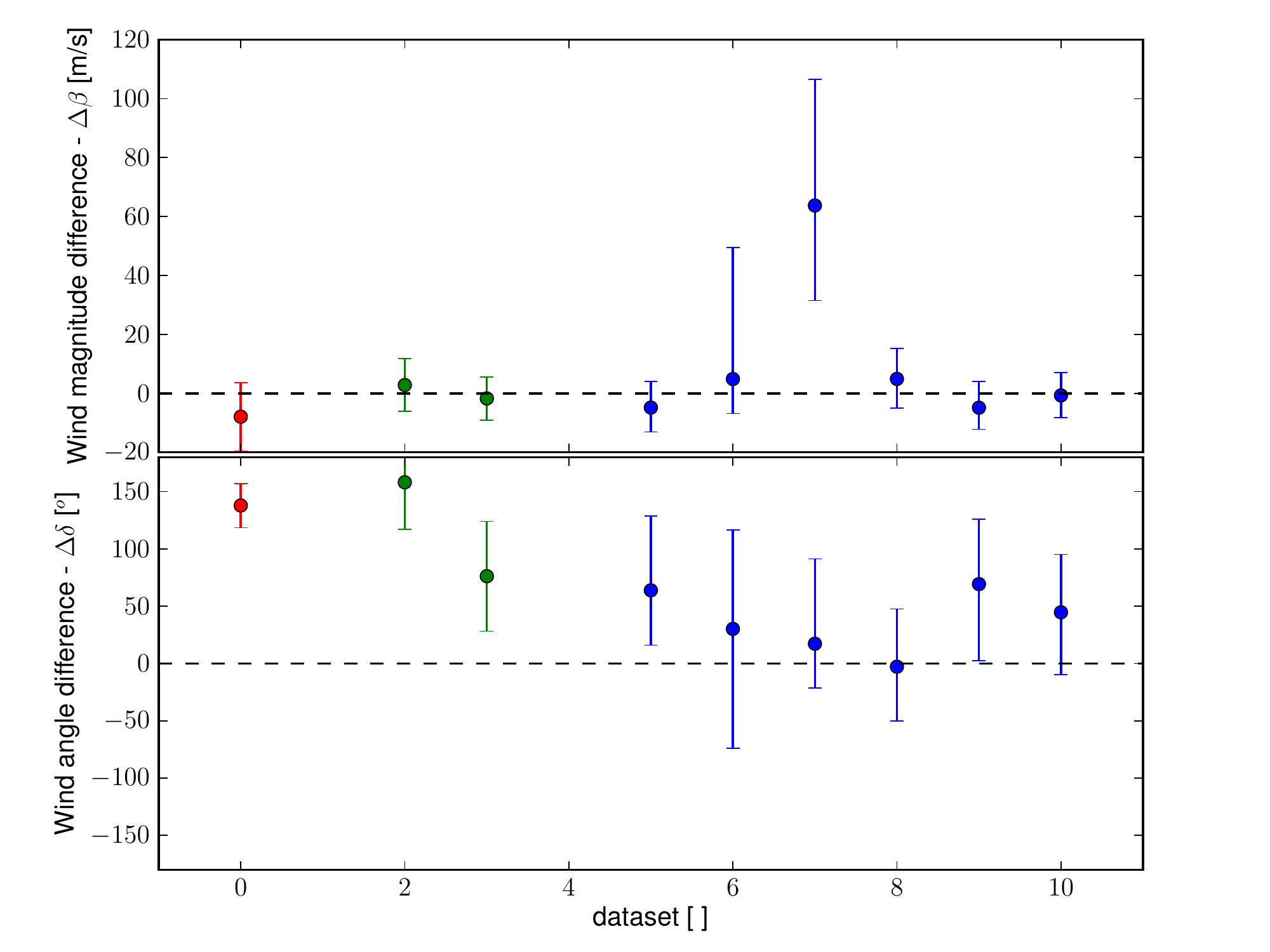}
\caption{Difference between the average wind magnitude ({\it upper panel}) and direction ({\it lower panel}) as measured by the two different methods. The values are presented in Tab. \ref{comparison}. The full dataset fit values are coded in red, those corresponding to single-night sets are coded in green, and the night subdivisions are coded in blue ({\it electronic version only}).}\label{comparison_plot}
\end{figure}

In order to compare the two datasets we computed the time center of the observations for each block and selected the probe measurement that was closest in time to it. Table \ref{comparison} displays the data in a way that allows an easy comparison of the average wind vector magnitude and direction, and the difference between the quantities obtained with the two different methods is presented in Fig.\,\ref{comparison_plot}. We considered for this purpose the unconstrained fit values, for they show higher $P_{KS}$. In what concerns wind magnitude the values from the probes agree with the fitted values from RV data, for all datasets. The only outlier is the third section of the first night, which presents a very large value of $\beta$ and strongly asymmetric error bars. As discussed before, the corresponding HARPS dataset has the largest photon noise contribution associated, and very poor azimuth coverage (as can be seen in Fig.\,\ref{altaz}) which can explain the lower quality of the fit. In terms of wind direction, the values concerning the fit of a subsection of the night agree with those derived from radiosonde measurements; those on a longer timescale do not. The most straightforward interpretation is that the constant horizontal wind hypothesis does not hold for large timescales. This is not surprising since wind vectors are variable over time. In other words, the fit provides a better description of the data than no fit, and the residuals are closer to gaussianity than the raw data, as seen, but the direction has no physical correspondence. However, and as stated before, one has to note that the  $\sigma$ and $ \sigma_{(O-C)}$ values are already quite close to the $\sigma_{ph}$ level, and that the ratio between any of the former and the latter is smaller than 2. Such ratio values are smaller than those obtained in \cite{2010A&A...515A.106F}, and we cannot discard the fact that we might be approaching the limit of extractable information from the current dataset. 

\begin{table*}
\caption{Weighted average of wind components $u$ and $v$, as  well as wind vector magnitude and direction for each of the radiosondes launch.}\label{comparison}

\begin{tabular}{cccccc} \hline\hline
 \ \  data set & $\beta$\,[m/s]  &  $\delta$\,[$^{o}$]  & \# probe (time distance) & $\|u + v\|$\,[m/s] & $\delta\,[^{\circ}]$ \\ \hline \hline
08+09-05-2009 		& 8.47$_{-0.68}^{+0.76}$ &   126.10$_{-5.56}^{+4.41}$  & 8 (4h)&    16.40 $\pm$ 11.57     &    -11.75 $\pm$ 18.49 \\[4pt]                        
08-05-2009 (1$^{st}$ n.) & 13.93$_{-2.61}^{+3.12}$ &  145.14$_{-13.00}^{+5.58}$ & 7 (1h) &    11.13 $\pm$ 8.50     &    -12.97 $\pm$ 38.69 \\[4pt]
09-05-2009 ($2^{nd}$ n.)    & 6.75$_{-0.50}^{+0.60}$ &   97.75$_{-7.91}^{+6.67}$  & 10 (1.5h) &    8.48 $\pm$ 7.28     &    21.58 $\pm$ 47.31 \\[4pt]
                                                                 
(1$^{st}$ n., section\,1/3) & 4.44$_{-0.89}^{+3.31}$ &  42.17$_{-17.60}^{+47.29}$ & 6 (1.5h) &    9.27 $\pm$ 8.22     &    -21.70 $\pm$ 44.49 \\[4pt]
                                                                 
(1$^{st}$ n., section\,2/3) & 15.99$_{-8.00}^{+43.88}$  & 17.22$_{-96.66}^{+77.04}$ & 7 (0.5h) &    11.13 $\pm$ 8.50     &    -12.97 $\pm$ 38.69\\[4pt]
                                                                 
(1$^{st}$ n., section\,3/3) & 74.86$_{-31.07}^{+41.90}$ & 4.28$_{-2.35}^{+63.12}$ & 7 (2.5h) &  11.13 $\pm$ 8.50     &    -12.97 $\pm$ 38.69\\[4pt]
                                                                 
(2$^{nd}$ n., section\,1/3) & 15.98$_{-3.19}^{+4.50}$ & 24.96$_{-7.08}^{+18.08}$ & 9 (1h) &    11.09 $\pm$ 9.37     &    27.74 $\pm$ 46.99\\[4pt]
                                                                 
(2$^{nd}$ n., section\,2/3) & 3.61$_{-0.78}^{+5.17}$ & 90.87$_{-47.05}^{+31.05}$  & 10 (1h) &    8.48 $\pm$ 7.28     &    21.58 $\pm$ 47.31 \\[4pt]
                                                                 
(2$^{nd}$ n., section\,3/3) & 7.78$_{-1.77}^{+2.77}$ & 66.26$_{-27.05}^{+17.98}$  & 10 (2h) &  8.48 $\pm$ 7.28     &    21.58 $\pm$ 47.31 \\[4pt]
 	
\hline \hline
\end{tabular}

\medskip

Note that $\delta$\,=\,0\,$^{o}$ when wind direction points towards North, and positive eastwards.

\end{table*}

It is arguable that the model might be over-simplistic, and the small number of parameters and observables might fundamentaly limit the RV signal it can reproduce. Some leads point in this direction, and we followed these to propose and test alternative models. However, no improvements were verified relative to the basic model presented before. We present the results of this rather more exploratory digression in Appendix A.

Perfect agreement between the HARPS observations and the radiosonde data can not be expected for a number of reasons listed below. The HARPS spectrum samples a pencil beam through the atmosphere when the star is being tracked, while the radiosonde performs in situ-measurements along its trajectory governed by the prevailing winds. Another drawback is that the atmosphere is only sampled up to an altitude of ~20\,km; however, at this altitude the density of O$_2$ is ten times lower than at the top of the mountain, so the weight $A_i$ is ten times smaller. Finally, the radiosonde is expected to oscillate like a pendulum in its ascent, introducing a signal in the measured RV which is not rooted in the wind vector it is intended to probe. 
In spite of all these limitations, the two data sets agree and provide a coherent picture of the atmospheric impact on RV variation, down to better than 5\,m/s and less than twice the estimated photon noise contribution on O$_2$ lines RV measurement.

A quantitative assessment of the stability of atmospheric absorption lines as presented here is of very practical value for astronomy. Telluric absorption features are imprinted on observations with astronomical spectrographs over a wide wavelength range, particularly in the infrared. On the one hand this constitutes a complication since the features overlay the spectrum of the astronomical target leading to blends and line shifts. Hence, all observations aiming at a high spectral fidelity in certain regions of the infra-red need to correct for atmospheric transmission \citep[e.g.][]{2010ApJ...723..684B, 2010A&A...524A..11S, 2011arXiv1103.0004M}, and not only necessarily in the context of RV measurements \citep[e.g.][]{2010SPIE.7735E.237U}. With a comprehensive characterization of the atmospheric stability, one can assess for the first time the impact of considering atmospheric lines to be at rest or characterized by a constant speed over a given period of time. Their stability (or lack thereof) can explain a fraction of the residuals obtained today when fitting the atmospheric transmission with a forward model, which yields residuals around a few \%. On the other hand the telluric features are also used for wavelength calibration again in particular in the infra-red where technical calibration sources are less common. For physical reasons atomic spectra emitted by lamps show fewer lines and a more uneven distribution than in the optical. The Th-Ar hollow cathode lamp is the only source whose spectrum has been fully characterized in the IR \citep{2008ApJS..178..374K}, and which is being used on CRIRES but there are limitations in line density and wavelength coverage. Gas cells usually also cover only a limited wavelength range. As a result telluric absorption lines are an attractive alternative in parts of the IR. Based on the results presented here the stability of the atmosphere will easily support low and medium-reslution spectroscopy, while in particular for high resolution and high precision work caution has to be applied. The actual wind velocity vector, and its variation, during the astronomical observations of course remains unknown without independent measurements. Hence, it is not possible to derive proper error bars for a quantitative analysis down to the m/s level unless a full analysis following the method described in \cite{2010A&A...515A.106F} is performed. 

\section{Conclusions}

We used HARPS to monitor the RV variation of O$_2$ lines in the optical wavelength domain. We compared the fitting of a model as described in \cite{2010A&A...515A.106F} and the obtained parameters with those delivered by contemporaneous radiosondes. The two approaches deliver the same results in what concerns wind magnitude and agree on wind direction when fitting is done in chunks of a couple of hours. The large uncertainty bars on the values obtained from radiosondes are likely to be a consequence of a complex wind structure as a function of height, a fact that weakens the applicability of the assumption of a strong horizontal wind. We cannot conclude if the $\alpha$ and $\gamma$ parameters should be constant as a function of time or not, or if a cross-term between them should be included. However, when these are fixed the wind direction does not agree with that extracted from the radiosonde, which suggests that the model might be incomplete at this level.  We tested two different alternative models that tried to address this possible incompleteness of the physical description, but the results were poorer than with the base model.

Statistical tests showed that the base model provides a good description of the data on all timescales, being always preferable to not fitting any atmospheric variation, and that the smaller the timescale on which it can be performed (down to a couple of hours), the better the description of the real physical parameters. It is important to note that it is for the datasets with higher $P_{KS}$ that the wind parameters derived from RV are compatible with those extracted from radiosondes measurement. Thus, even though the model presented in \cite{2010A&A...515A.106F} can probably still be refined, the agreement is proven down to better than 5\,m/s and less than twice the estimated photon noise contribution on O$_2$ lines RV measurement.

\section*{Acknowledgments}
This work was supported by the European Research Council/European Community under the FP7 through Starting Grant agreement
number 239953, as well as by Funda\c{c}\~ao para a Ci\^encia e a Tecnologia (FCT) in the form of grant reference
PTDC/CTE-AST/098528/2008. NCS would further like to thank FCT through program Ci\^encia\,2007 funded by FCT/MCTES (Portugal) and POPH/FSE (EC).

This work has been funded by the E-ELT project in the context of site characterization. The measurements have been made possible by the coordinated effort of the PWV project team and the observatory site hosting it. We would like to thank all the technical staff, astronomers and telescope operators at La Silla who have helped us in setting up equipment, operating instruments and supporting parallel observations. We thank the Directors of La Silla Paranal observatory (A. Kaufer, M. Sterzik, U. Weilenmann) for accommodating such a demanding project in the operational environment of the observatory and for granting technical time. We particularly thank the ESO chief representative in Chile, M. Tarenghi for his support.  It is a pleasure to thank the Chilean Direction General de Aeronautica Civil (in particular J. Sanchez) for the helpful collaboration and for reserving airspace around the observatories to ensure a safe environment for launches of radiosonde balloons. Special thanks to the members of the Astrometeorological Group at the Universidad de Valparaiso who supported the radiosonde launches. 

\bibliographystyle{mn2e} 
\bibliography{Mybibliog_MNRAS} 

\begin{thebibliography}{}

\bibitem[\protect\citeauthoryear{{Andrae}, {Schulze-Hartung} \&
  {Melchior}}{{Andrae} et~al.}{2010}]{2010arXiv1012.3754A}
{Andrae} R.,  {Schulze-Hartung} T.,    {Melchior} P.,  2010, ArXiv e-prints

\bibitem[\protect\citeauthoryear{{Baranne}, {Queloz}, {Mayor}, {Adrianzyk},
  {Knispel}, {Kohler}, {Lacroix}, {Meunier}, {Rimbaud} \& {Vin}}{{Baranne}
  et~al.}{1996}]{1996A&AS..119..373B}
{Baranne} A.,  {Queloz} D.,  {Mayor} M.,  {Adrianzyk} G.,  {Knispel} G.,
  {Kohler} D.,  {Lacroix} D.,  {Meunier} J.-P.,  {Rimbaud} G.,    {Vin} A.,
  1996, \aaps, 119, 373

\bibitem[\protect\citeauthoryear{{Blake}, {Charbonneau} \& {White}}{{Blake}
  et~al.}{2010}]{2010ApJ...723..684B}
{Blake} C.~H.,  {Charbonneau} D.,    {White} R.~J.,  2010, \apj, 723, 684

\bibitem[\protect\citeauthoryear{{Butler}, {Marcy}, {Williams}, {McCarthy},
  {Dosanjh} \& {Vogt}}{{Butler} et~al.}{1996}]{1996PASP..108..500B}
{Butler} R.~P.,  {Marcy} G.~W.,  {Williams} E.,  {McCarthy} C.,  {Dosanjh} P.,
    {Vogt} S.~S.,  1996, \pasp, 108, 500

\bibitem[\protect\citeauthoryear{{Chac{\'o}n}, {Cuevas}, {Pozo},
  {Mar{\'{\i}}n}, {Oyanadel}, {Dougnac}, {Cortes}, {Illanes}, {Caneo},
  {Cur{\'e}}, {Sarazin}, {Kerber}, {Smette}, {Rabanus}, {Querel} \&
  {Tompkins}}{{Chac{\'o}n} et~al.}{2010}]{2010SPIE.7733E.143C}
{Chac{\'o}n} A.,  {Cuevas} O.,  {Pozo} D.,  {Mar{\'{\i}}n} J.,  {Oyanadel} A.,
  {Dougnac} C.,  {Cortes} L.,  {Illanes} L.,  {Caneo} M.,  {Cur{\'e}} M.,
  {Sarazin} M.,  {Kerber} F.,  {Smette} A.,  {Rabanus} D.,  {Querel} R.,
  {Tompkins} G.,  2010, in Society of Photo-Optical Instrumentation Engineers
  (SPIE) Conference Series Vol.~7733, {Measuring and forecasting of PWV above
  La Silla, APEX and Paranal Observatories}

\bibitem[\protect\citeauthoryear{{Figueira}, {Pepe}, {Lovis} \&
  {Mayor}}{{Figueira} et~al.}{2010}]{2010A&A...515A.106F}
{Figueira} P.,  {Pepe} F.,  {Lovis} C.,    {Mayor} M.,  2010, \aap, 515, A106+

\bibitem[\protect\citeauthoryear{{Figueira}, {Pepe}, {Melo}, {Santos}, {Lovis},
  {Mayor}, {Queloz}, {Smette} \& {Udry}}{{Figueira}
  et~al.}{2010}]{2010A&A...511A..55F}
{Figueira} P.,  {Pepe} F.,  {Melo} C.~H.~F.,  {Santos} N.~C.,  {Lovis} C.,
  {Mayor} M.,  {Queloz} D.,  {Smette} A.,    {Udry} S.,  2010, \aap, 511, A55+

\bibitem[\protect\citeauthoryear{{Holton}}{{Holton}}{2005}]{Holton2005}
{Holton} J.,  2005, {An Introduction to Dynamic Meteorology}

\bibitem[\protect\citeauthoryear{{Kaeufl}, {Ballester}, {Biereichel},
  {Delabre}, {Donaldson}, {Dorn}, {Fedrigo}, {Finger}, {Fischer} \&
  {Franza}}{{Kaeufl} et~al.}{2004}]{2004SPIE.5492.1218K}
{Kaeufl} H.-U.,  {Ballester} P.,  {Biereichel} P.,  {Delabre} B.,  {Donaldson}
  R.,  {Dorn} R.,  {Fedrigo} E.,  {Finger} G.,  {Fischer} G.,    {Franza} F.
  e.~a.,  2004, in {Moorwood} A.~F.~M.,  {Iye} M.,  eds, Society of
  Photo-Optical Instrumentation Engineers (SPIE) Conference Series Vol.~5492 of
  Society of Photo-Optical Instrumentation Engineers (SPIE) Conference Series,
  {CRIRES: a high-resolution infrared spectrograph for ESO's VLT}.
pp 1218--1227

\bibitem[\protect\citeauthoryear{{Kalthoff}, {Bischoff-Gau{\ss}},
  {Fiebig-Wittmaack} \& {Fiedler}}{{Kalthoff}
  et~al.}{2002}]{Norbert-Kalthoff:2002cr}
{Kalthoff} N.,  {Bischoff-Gau{\ss}} I.,  {Fiebig-Wittmaack} M.,    {Fiedler} F.
  e.~a.,  2002, Journal of Applyed Meteorology, 41, 953

\bibitem[\protect\citeauthoryear{{Kerber}, {Nave} \& {Sansonetti}}{{Kerber}
  et~al.}{2008}]{2008ApJS..178..374K}
{Kerber} F.,  {Nave} G.,    {Sansonetti} C.~J.,  2008, \apjs, 178, 374

\bibitem[\protect\citeauthoryear{{Kerber}, {Querel}, {Hanuschik}, {Chac{\'o}n},
  {Caneo}, {Cortes}, {Cure}, {Illanes}, {Naylor}, {Smette}, {Sarazin},
  {Rabanus} \& {Tompkins}}{{Kerber} et~al.}{2010}]{2010SPIE.7733E..48K}
{Kerber} F.,  {Querel} R.~R.,  {Hanuschik} R.~W.,  {Chac{\'o}n} A.,  {Caneo}
  M.,  {Cortes} L.,  {Cure} M.,  {Illanes} L.,  {Naylor} D.~A.,  {Smette} A.,
  {Sarazin} M.,  {Rabanus} D.,    {Tompkins} G.,  2010, in Society of
  Photo-Optical Instrumentation Engineers (SPIE) Conference Series Vol.~7733,
  {Support for site testing of the European Extremely Large Telescope:
  precipitable water vapor over Paranal}

\bibitem[\protect\citeauthoryear{{Mayor}, {Pepe}, {Queloz}, {Bouchy},
  {Rupprecht}, {Lo Curto}, {Avila}, {Benz}, {Bertaux} \& {Bonfils}}{{Mayor}
  et~al.}{2003}]{2003Msngr.114...20M}
{Mayor} M.,  {Pepe} F.,  {Queloz} D.,  {Bouchy} F.,  {Rupprecht} G.,  {Lo
  Curto} G.,  {Avila} G.,  {Benz} W.,  {Bertaux} J.-L.,    {Bonfils} X. e.~a.,
  2003, The Messenger, 114, 20

\bibitem[\protect\citeauthoryear{{Mayor} \& {Queloz}}{{Mayor} \&
  {Queloz}}{1995}]{1995Natur.378..355M}
{Mayor} M.,  {Queloz} D.,  1995, \nat, 378, 355

\bibitem[\protect\citeauthoryear{{Miloshevich}, {Vomel}, {Whiteman} \&
  {LeBlanc}}{{Miloshevich} et~al.}{2009}]{Milosh2009}
{Miloshevich} L.~M.,  {Vomel} H.,  {Whiteman} D.~N.,    {LeBlanc} T.,  2009, J.
  Geophys Res., p.~114

\bibitem[\protect\citeauthoryear{{Muirhead}, {Edelstein}, {Erskine}, {Wright},
  {Muterspaugh}, {Covey}, {Wishnow}, {Hamren}, {Andelson}, {Kimber}, {Mercer},
  {Halverson}, {Vanderburg}, {Mondo}, {Czeszumska} \& {Lloyd}}{{Muirhead}
  et~al.}{2011}]{2011arXiv1103.0004M}
{Muirhead} P.~S.,  {Edelstein} J.,  {Erskine} D.~J.,  {Wright} J.~T.,
  {Muterspaugh} M.~W.,  {Covey} K.~R.,  {Wishnow} E.~H.,  {Hamren} K.,
  {Andelson} P.,  {Kimber} D.,  {Mercer} T.,  {Halverson} S.,  {Vanderburg} A.,
   {Mondo} D.,  {Czeszumska} A.,    {Lloyd} J.~P.,  2011, ArXiv e-prints

\bibitem[\protect\citeauthoryear{{Pepe}, {Mayor}, {Galland}, {Naef}, {Queloz},
  {Santos}, {Udry} \& {Burnet}}{{Pepe} et~al.}{2002}]{2002A&A...388..632P}
{Pepe} F.,  {Mayor} M.,  {Galland} F.,  {Naef} D.,  {Queloz} D.,  {Santos}
  N.~C.,  {Udry} S.,    {Burnet} M.,  2002, \aap, 388, 632

\bibitem[\protect\citeauthoryear{{Press}, {Teukolsky}, {Vetterling} \&
  {Flannery}}{{Press} et~al.}{1992}]{1992nrfa.book.....P}
{Press} W.~H.,  {Teukolsky} S.~A.,  {Vetterling} W.~T.,    {Flannery} B.~P.,
  1992, {Numerical recipes in FORTRAN. The art of scientific computing}

\bibitem[\protect\citeauthoryear{{Querel}, {Kerber}, {Lo Curto}, {Thomas-Osip},
  {Prieto}, {Chac{\'o}n}, {Cuevas}, {Pozo}, {Mar{\'{\i}}n}, {Naylor},
  {Cur{\'e}}, {Sarazin}, {Guirao} \& {Avila}}{{Querel}
  et~al.}{2010}]{2010SPIE.7733E.135Q}
{Querel} R.~R.,  {Kerber} F.,  {Lo Curto} G.,  {Thomas-Osip} J.~E.,  {Prieto}
  G.,  {Chac{\'o}n} A.,  {Cuevas} O.,  {Pozo} D.,  {Mar{\'{\i}}n} J.,  {Naylor}
  D.~A.,  {Cur{\'e}} M.,  {Sarazin} M.~S.,  {Guirao} C.,    {Avila} G.,  2010,
  in Society of Photo-Optical Instrumentation Engineers (SPIE) Conference
  Series Vol.~7733, {Support for site testing of the European Extremely Large
  Telescope: precipitable water vapor over La Silla}

\bibitem[\protect\citeauthoryear{{Querel}, {Naylor} \& {Kerber}}{{Querel}
  et~al.}{2011}]{2011PASP..123..222Q}
{Querel} R.~R.,  {Naylor} D.~A.,    {Kerber} F.,  2011, \pasp, 123, 222

\bibitem[\protect\citeauthoryear{{Rothman}, {Jacquemart}, {Barbe}, {Chris
  Benner}, {Birk}, {Brown}, {Carleer}, {Chackerian}, {Chance} \& et
  al.}{{Rothman} et~al.}{2005}]{2005JQSRT..96..139R}
{Rothman} L.~S.,  {Jacquemart} D.,  {Barbe} A.,  {Chris Benner} D.,  {Birk} M.,
   {Brown} L.~R.,  {Carleer} M.~R.,  {Chackerian} C.,  {Chance} K.,    et al.
  C.,  2005, Journal of Quantitative Spectroscopy and Radiative Transfer, 96,
  139

\bibitem[\protect\citeauthoryear{{Seifahrt}, {K{\"a}ufl}, {Z{\"a}ngl}, {Bean},
  {Richter} \& {Siebenmorgen}}{{Seifahrt} et~al.}{2010}]{2010A&A...524A..11S}
{Seifahrt} A.,  {K{\"a}ufl} H.~U.,  {Z{\"a}ngl} G.,  {Bean} J.~L.,  {Richter}
  M.~J.,    {Siebenmorgen} R.,  2010, \aap, 524, A11+

\bibitem[\protect\citeauthoryear{{Thomas-Osip}, {McWilliam}, {Phillips},
  {Morrell}, {Thompson}, {Folkers}, {Adams} \& {Lopez-Morales}}{{Thomas-Osip}
  et~al.}{2007}]{2007PASP..119..697T}
{Thomas-Osip} J.,  {McWilliam} A.,  {Phillips} M.~M.,  {Morrell} N.,
  {Thompson} I.,  {Folkers} T.,  {Adams} F.~C.,    {Lopez-Morales} M.,  2007,
  \pasp, 119, 697

\bibitem[\protect\citeauthoryear{{Uttenthaler}, {Pontoppidan}, {Seifahrt},
  {Kendrew}, {Blommaert}, {Pantin}, {Brandl}, {Molster}, {Venema}, {Lenzen},
  {Parr-Burman} \& {Siebenmorgen}}{{Uttenthaler}
  et~al.}{2010}]{2010SPIE.7735E.237U}
{Uttenthaler} S.,  {Pontoppidan} K.~M.,  {Seifahrt} A.,  {Kendrew} S.,
  {Blommaert} J.~A.~D.~L.,  {Pantin} E.~J.,  {Brandl} B.~R.,  {Molster} F.~J.,
  {Venema} L.,  {Lenzen} R.,  {Parr-Burman} P.,    {Siebenmorgen} R.,  2010, in
  Society of Photo-Optical Instrumentation Engineers (SPIE) Conference Series
  Vol.~7735, {Correcting METIS spectra for telluric absorption to maximize
  spectral fidelity}

\bibitem[\protect\citeauthoryear{{Wallace} \& {Hobbs}}{{Wallace} \&
  {Hobbs}}{2006}]{Wallace2006}
{Wallace} J.,  {Hobbs} P.,  2006, {Atmospheric Science, an introductory Survey}

\end{thebibliography}

\appendix

\section{Improving the model}

The global picture obtained by analyzing the datasets first separately and then together allows one to raise some interesting questions; within this questions is the potential for improving the model.
The first point to note is the impact on the error bars of the fitted parameters when splitting the RV data in subsets for the fitting. When separating the data set in two nights, the average error bars for each parameter increase by a factor which can be slightly in excess of $\sqrt{2}$ (depending on the case, for some the increase is much more modest), but when the nights are divided into subsets, the increase in the error bars exceeds that expected from the reduction of the number of data points used for the fit. In particular, the relative increase for the $\delta$ error bars is much larger than those for the other parameters. Another interesting point comes then into view: the second and third sections of the first night show comparatively large error bars for the four parameters. However, this point is to be taken carefully because, as discussed, the photon noise was higher and the azimuth coverage not as complete as for the other datasets. 

These two elements point towards a cross-talk between the model parameters. Given the simplicity of the model and large number of data points available, it is more likely that this behavior comes from trying to fit a too simple model rather than being caused by lack of conditioning.

In addition, a poorer match between wind direction from the two methods for the constrained fit suggests that $\alpha$ and $\gamma$ might not be constant for the datasets at hand. This is particularly clear for $\alpha$, while variations on $\gamma$ are only of a couple of m/s. Such a behavior can be explained by a correlation between $\alpha$ and a parameter which represents a quantity expected to change with time. It can also be explained if this coefficient has an intrinsic dependence on time. And, naturally, it can also be explained by a dependence of the model parameters -- or even the RV itself -- on a single (unrepresented) quantity.

The most important hint is probably the high variability of $\delta$ and large error bars on its determination: this suggests that either the variation associated with this coefficient is defined in an incomplete fashion or that some other quantity has a similar functional dependence on the parameter which the $\delta$ variation tries to accommodate.

When this information is put together one concludes that the most likely improvement to the model is an additional dependence of the airmass impact on RV on the direction of the wind. This is not completely unexpected as a consequence of the chosen model parametrization.

In \cite{2010A&A...515A.106F} we had already suggested that if the atmosphere has a complex vertical wind structure which cannot be represented by a single average wind value, $\alpha$ might not be considered as constant. It is so because an increased broadening of the CCF (due to the span of velocities that displace the absorber) will change the correlation coefficient between the broadening and the impact on the RV. As a consequence it will change the coefficient between airmass and RV, our $\alpha$. To fully characterize the impact of this wind broadening contribution to the $\alpha$ coefficient is extremely difficult and requires a line-formation model of the atmosphere, which is beyond the scope of this work. However, we can propose a refinement of our model in order to include this effect, and we test it tentatively in two alternative parametrizations to Eq.\,\ref{eq_fit}:

\begin{equation}
	\Omega' =  \left[ \alpha \times  {\left({1\over sin(\theta)} - 1\right)} + \beta \right] \times cos(\theta) \,.\, cos(\phi -  \delta) + \gamma \label{eq_fit_alt1}
\end{equation}

\begin{equation}
	\Omega'' =  \left[ \alpha \times  {\left({1\over sin(\theta)} - 1\right)} + \beta \,.\, cos(\theta) \right] \times cos(\phi -  \delta) + \gamma \label{eq_fit_alt2}
\end{equation}

In Eq.\ref{eq_fit_alt1} we consider $\alpha$ to be dependent on the colinearity with the wind direction. This is expected to be the case if there is a scatter of velocity around the central velocity $\beta$. In this parametrization $\alpha$ contains two components: the dependence on airmass and the broadening created by the scatter in velocity associated with it. In Eq. \ref{eq_fit_alt2} we consider a variation on this assumption in which only the wind direction (and not the projection of this direction along the line of sight) has an impact on the measured RV. The fitted parameters and quantities associated with each dataset are presented in Tab. \ref{fitstats_alt}. The results of the aplication of the Kolmogorov-Smirnov test and the P$_{KS}$ derived for the two cases are presented in Tab. \ref{KS_prob_alt}. 

\begin{table*}

\caption{The fitted parameters and data properties, before and after the fitted model is subtracted from it, when Eq. \ref{eq_fit_alt1} ({\it top}), and Eq. \ref{eq_fit_alt2} ({\it bottom}) are considered.}\label{fitstats_alt}

\begin{tabular}{lccccccccc} \hline\hline

 \ \  data set  &  $\#$obs  & $\sigma$\,[m/s] &  $\sigma_{(O-C)}$\,[m/s]  &  $\sigma_{ph}$\,[m/s] & $\chi^{2}_{red}$  & $\alpha$\,[m/s] &  $\beta$\,[m/s]  & $\gamma$\,[m/s] &  $\delta$\,[$^{o}$] \\[1pt] \hline \hline


08+09-05-2009 & 1093 &   5.01 &   4.25 &    2.82  &   2.35  &  22.67$_{-5.58}^{+5.34}$ &    7.16$_{-1.93}^{+1.98}$ & 220.51$_{-0.49}^{+0.19}$ & 151.66$_{-2.26}^{+2.25}$ \\[2pt]
\hline

08-05-2009 & 554 &   5.36 &   4.39 &    2.92  &   2.16  &  12.39$_{-7.81}^{+7.62}$ &   15.41$_{-4.31}^{+4.62}$ & 219.89$_{-1.07}^{+0.30}$& 156.35$_{-3.04}^{+3.30}$ \\[4pt]
09-05-2009 & 539 &   4.60 &   3.86 &    2.72  &   2.09  &  26.96$_{-9.82}^{+7.52}$ &    3.91$_{-2.38}^{+2.83}$ & 219.94$_{-0.01}^{+0.84}$ & 145.13$_{-4.94}^{+3.29}$ \\[2pt]

\hline

 (1$^{st}$ n., section\,1/3)  & 185 &   3.46 &   2.95 &    1.88  &   2.41  &  18.02$_{-8.11}^{+4.62}$ &    2.47$_{-2.47}^{+5.51}$ & 222.22$_{-0.75}^{+1.51}$ & 159.31$_{-3.64}^{+20.49}$ \\[4pt]
 (1$^{st}$ n., section\,2/3)  & 185 &   4.58 &   4.58 &    3.27  &   1.71  &  -9.49$_{-211.15}^{+194.11}$ &    8.30$_{-8.30}^{+97.43}$ & 222.94$_{-13.76}^{+3.62}$ &  24.38$_{-15.99}^{+306.33}$ \\[4pt]
 (1$^{st}$ n., section\,3/3)  & 184 &   6.26 &   4.52 &    3.60  &   1.54  & -25.62$_{-129.03}^{+33.19}$ &   24.41$_{-24.41}^{+34.99}$ & 225.04$_{-9.71}^{+4.60}$&  27.60$_{-18.90}^{+151.36}$ \\[4pt]

 (2$^{nd}$ n., section\,1/3)  & 180 &   4.80 &   3.87 &    2.52  &   2.36  &  18.43$_{-8.38}^{+7.45}$ &   16.77$_{-16.77}^{+5.29}$ & 216.41$_{-0.85}^{+4.34}$ & 152.53$_{-2.60}^{+18.40}$ \\[4pt]
 (2$^{nd}$ n., section\,2/3)  & 180 &   4.50 &   3.74 &    2.84  &   1.78  &  11.72$_{-75.87}^{+92.59}$ &    2.96$_{-2.96}^{+5.99}$ & 220.56$_{-0.51}^{+1.49}$ &  71.21$_{-55.11}^{+116.77}$ \\[4pt]
 (2$^{nd}$ n., section\,3/3)  & 180 &   3.86 &   3.85 &    2.82  &   1.91  & -11.42$_{-154.65}^{+140.58}$ &    0.00$_{-0.00}^{+7.31}$ & 219.11$_{-0.85}^{+1.74}$ & 179.39$_{-107.80}^{+58.91}$ \\[2pt]

 \hline \hline

08+09-05-2009 & 1093 &   5.01 &   4.26 &    2.82  &   2.37  &  20.23$_{-5.12}^{+4.91}$ &    4.95$_{-2.36}^{+2.37}$ & 220.53$_{-0.48}^{+0.20}$ & 151.58$_{-2.38}^{+2.26}$ \\[2pt]

\hline

08-05-2009 & 554 &   5.36 &   4.39 &    2.92  &   2.18  &  10.54$_{-7.39}^{+6.98}$ &   14.53$_{-4.85}^{+5.59}$ & 219.87$_{-1.11}^{+0.31}$ &156.45$_{-3.14}^{+3.30}$ \\[4pt]
09-05-2009 & 539 &   4.60 &   3.86 &    2.72  &   2.09  &  23.80$_{-8.87}^{+5.73}$ &    1.40$_{-1.40}^{+3.67}$ & 219.97$_{-0.01}^{+0.84}$ & 144.63$_{-5.13}^{+3.15}$ \\[2pt]

\hline

 (1$^{st}$ n., section\,1/3)  & 185 &   3.46 &   2.98 &    1.88  &   2.48  &  16.13$_{-6.94}^{+3.51}$ &    0.68$_{-0.68}^{+6.20}$ & 222.22$_{-0.84}^{+1.27}$ & 159.69$_{-3.70}^{+15.13}$ \\[4pt]
 (1$^{st}$ n., section\,2/3)  & 185 &   4.58 &   4.59 &    3.27  &   1.71  & -48.80$_{-164.41}^{+148.22}$ &   16.10$_{-16.10}^{+104.09}$ & 220.82$_{-12.73}^{+7.64}$ & 153.95$_{-146.59}^{+173.86}$ \\[4pt]
 (1$^{st}$ n., section\,3/3)  & 184 &   6.26 &   4.52 &    3.60  &   1.54  & -25.90$_{-84.33}^{+32.05}$ &   29.80$_{-29.80}^{+39.14}$ & 225.28$_{-10.44}^{+3.96}$ &  25.39$_{-15.85}^{+152.00}$ \\[4pt]
 (2$^{nd}$ n., section\,1/3)  & 180 &   4.80 &   3.91 &    2.52  &   2.41  &  15.60$_{-8.13}^{+6.85}$ &   15.36$_{-15.36}^{+5.78}$ & 216.40$_{-0.86}^{+4.60}$ & 152.52$_{-2.54}^{+20.44}$ \\[4pt]
 (2$^{nd}$ n., section\,2/3)  & 180 &   4.50 &   3.75 &    2.84  &   1.78  &  10.24$_{-39.05}^{+52.66}$ &    2.00$_{-2.00}^{+7.20}$ & 220.43$_{-0.56}^{+1.74}$ &  91.03$_{-67.21}^{+84.63}$ \\[4pt]
 (2$^{nd}$ n., section\,3/3)  & 180 &   3.86 &   3.83 &    2.82  &   1.89  &  -9.03$_{-54.32}^{+45.37}$ &    0.00$_{-0.00}^{+6.91}$ & 219.01$_{-1.10}^{+1.87}$ & 179.46$_{-101.71}^{+68.26}$ \\[2pt]

\hline \hline

\end{tabular}

\medskip

Note that $\delta$\,=\,0\,$^{o}$ when wind direction points towards North, and positive eastwards. The error bars on each of the fitted parameters were drawn by bootstrapping the residuals (see text for details). 

\end{table*}

\begin{table*}
\centering
\begin{tabular}{lccc} \hline\hline
 \ \ data set & P$_{KS}(\Omega')$ & P$_{KS}(\Omega'')$  &  P$_{KS}$(no fit)  \\ \hline \hline

08+09-05-2009   & 2.51e-11      &       6.74e-11        & 4.32e-25 \\ 

08-05-2009         & 7.64e-07      &       3.19e-07        & 6.28e-19 \\ 

09-05-2009        & 3.08e-06      &       3.51e-06        & 2.62e-10 \\ 

 (1$^{st}$ n., section\,1/3)          & 1.15e-02      &       8.14e-03        & 1.22e-05 \\ 

 (1$^{st}$ n., section\,2/3)      & 4.17e-02      &       3.22e-02        & 3.02e-02 \\ 

 (1$^{st}$ n., section\,3/3)        & 1.66e-01      &       1.84e-01        & 1.07e-05 \\ 

 (2$^{nd}$ n., section\,1/3)        & 1.09e-03      &       1.36e-03        & 5.79e-08 \\ 

 (2$^{nd}$ n., section\,2/3)         & 1.03e-01      &       9.67e-02        & 4.66e-03 \\ 

 (2$^{nd}$ n., section\,3/3)        & 2.95e-02      &       4.80e-02        & 5.33e-02 \\ 

\hline \hline
\end{tabular}

\caption{The probability that the residuals and original datasets are drawn from gaussian distributions, as estimated used Kolmogorov-Smirnov test (see text for details), now for the alternative models represented by Eq. \ref{eq_fit_alt1} (P$_{KS}(\Omega')$), and Eq. \ref{eq_fit_alt2} (P$_{KS}(\Omega'')$). } \label{KS_prob_alt}
\end{table*}

Unfortunately, these modifications do not lead to an improvement. The P$_{KS}$ are smaller than in the previously considered cases (and the $\chi^{2}_{red}$ are larger). The cross-talk between the different parameters is increased, with the $\beta$ parameter reaching zero within the 1-$\sigma$ uncertainties (given the way they were calculated this only means that a large number of datasets of the MC was best fitted by $\beta$=0.) As a consequence, one is forced to conclude that these alternative models increase the correlation or cross-talk between parameters, instead of reducing it.

One can conceive a model in which the dependence on altitude and azimuth is concentrated in the parameters in a different way, but this dependence should stem from a physical motivation. We are then led to conclude that we probably reached the limit of extractable information from this dataset and an improvement on the quality of the measurements is required to take this kind of analysis any further.

\label{lastpage}

\end{document}